\documentclass[useAMS,usedcolumn,usenatbib]{mn2e}

\usepackage{epsfig,graphicx,natbib,times}
\usepackage{./reference/mycite}
\newcommand{\Ha}{\ifmmode {\rm H}\alpha \else H$\alpha$\fi}
\newcommand{\Hb}{\ifmmode {\rm H}\beta \else H$\beta$\fi}
\newcommand{\Nii}{[N~{\sc ii}]$\lambda$6584}
\newcommand{\nii}{\ifmmode [\rm{N}\,\textsc{ii}] \else [N~{\sc ii}]\fi}
\newcommand{\Oi}{[O~{\sc i}]$\lambda$6300}
\newcommand{\oi}{\ifmmode [\rm{O}\,\textsc{i}] \else [O~{\sc i}]\fi}
\newcommand{\Oii}{[O~{\sc ii}]$\lambda$3727}
\newcommand{\oii}{\ifmmode [\rm{O}\,\textsc{ii}] \else [O~{\sc ii}]\fi}
\newcommand{\Oiii}{[O~{\sc iii}]$\lambda$5007}
\newcommand{\oiii}{\ifmmode [\rm{O}\,\textsc{iii}] \else [O~{\sc iii}]\fi}

\begin{document}
\title[Spectral features from broad-band photometry]{Predicting spectral features in galaxy spectra from broad-band photometry}
\author[]
{F. B. Abdalla$^{1}$,
A. Mateus$^{2}$,
W. A. Santos$^{3}$,
L. Sodr$\acute{\rm e}$ Jr.$^{3}$,
I. Ferreras$^{4}$,
O. Lahav$^{1}$\\
$^{1}$Department of Physics and Astronomy, University College London,
Gower Street, London, WC1E 6BT, UK.\\
$^{2}$Instituto de Ciencias del Espacio (IEEC-CSIC), Barcelona, Spain. \\
$^{3}$Departamento de Astronomia, IAG-USP, Rua do Mat{\~a}o 1226, 05508-090, Sao Paulo, Brazil.\\
$^{4}$Mullard Space Science Laboratory, University College London,
Holmbury St Mary, Dorking, RH5 6NT, UK.
}
\maketitle

\begin{abstract}

We explore the prospects of predicting emission line features present
in galaxy spectra given broad-band photometry alone.  There is a
general consent that colours, and spectral features, most notably the
4000 \AA\ break, can predict many properties of galaxies, including
star formation rates and hence they could infer some of the line
properties.  We argue that these techniques have great prospects in
helping us understand line emission in extragalactic objects and might
speed up future galaxy redshift surveys if they are to target emission
line objects only.  We use two independent methods, Artifical Neural
Neworks (based on the ANNz code) and 
Locally Weighted Regression
(LWR), to retrieve correlations present in the colour N-dimensional
space and to predict the equivalent widths present in the corresponding
spectra. We also investigate how well it is possible to separate
galaxies with and without lines from broad band photometry only. We
find, unsurprisingly, that recombination lines can be well predicted
by galaxy colours.  However, among collisional lines some can and some
cannot be predicted well from galaxy colours alone, without any
further redshift information. We also use our techniques to estimate
how much information contained in spectral diagnostic diagrams can be
recovered from broad-band photometry alone.  We find that it is
possible to classify AGN and star formation objects relatively well
using colours only. We suggest that this technique could be used to
considerably improve redshift surveys such as the upcoming FMOS survey
and the planned WFMOS survey.

\end{abstract}


\section{Introduction}

\begin{figure*}
\begin{center}
\includegraphics[width=8.5cm,angle=0]{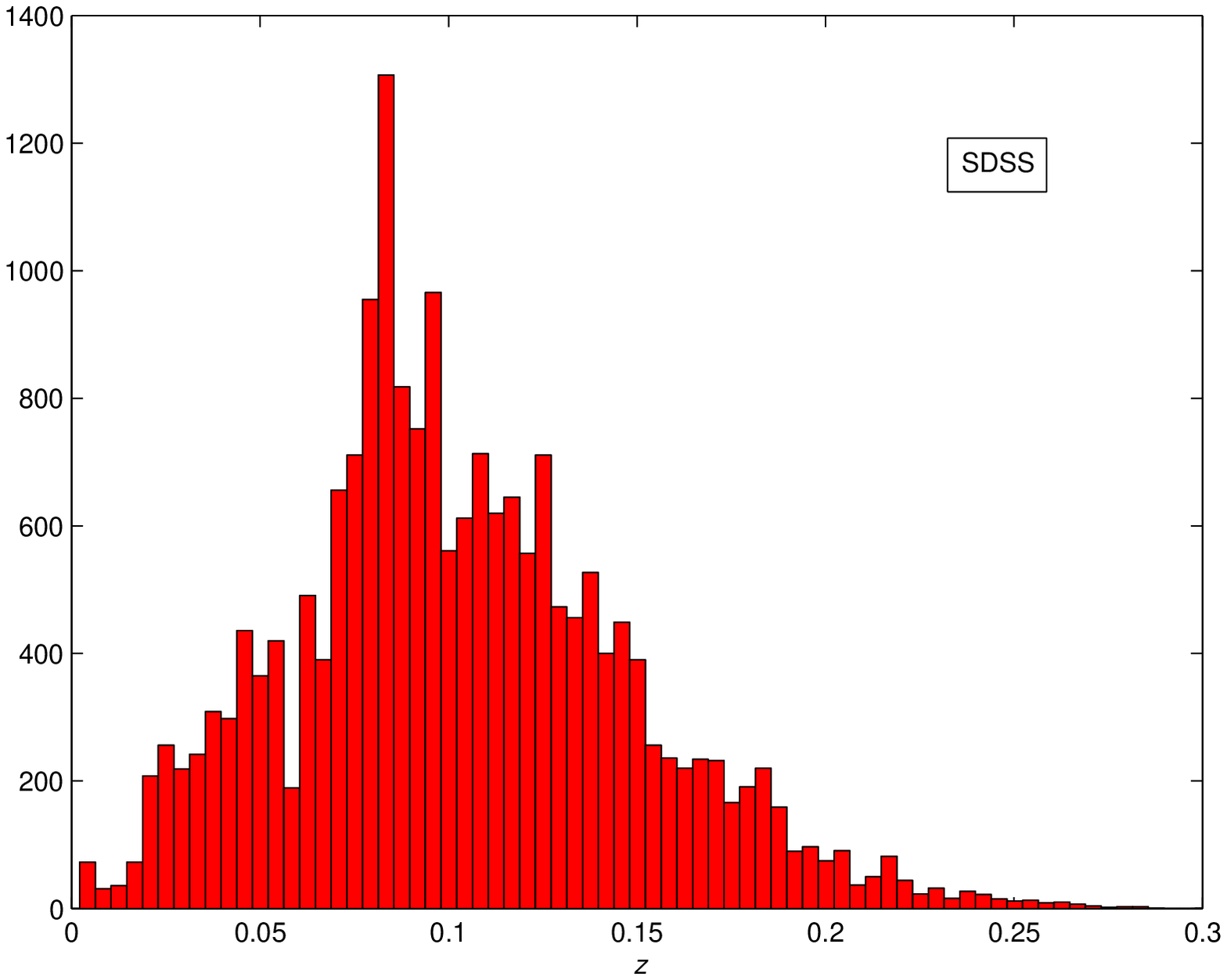}
\includegraphics[width=8.5cm,angle=0]{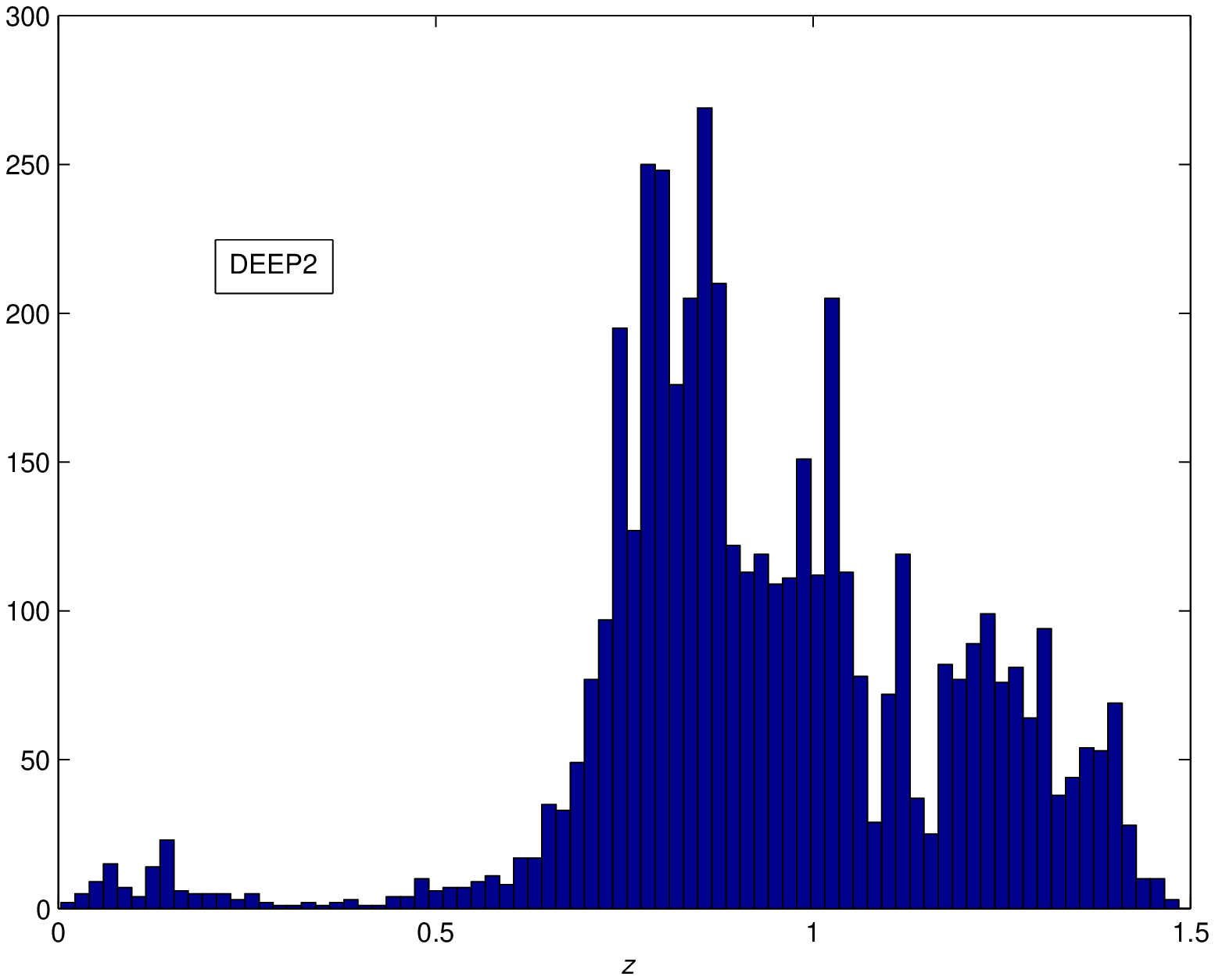}
\caption{The redshift distribution for both of the samples considered in 
this work. The median redshifts of SDSS and DEEP2 are  $z \sim 0.1$ 
and $1.0$ respectively. \label{fig:nz}}
\end{center}
\end{figure*}

Current galaxy formation models can be very successful in predicting 
the general spectral energy distribution (SED) of 
galaxies at moderate redshifts
by evolving stellar population models and co-adding a certain number of 
stellar populations with different ages \citep[e.g.][]{2004A&A...425..881L}. 
These models have become mainstream in obtaining
stellar population properties (like ages) 
given current data. With them, constraints can be imposed 
on the formation times of old elliptical galaxies, the epoch of 
reionization and cosmology 
\citep[e.g.][]{2002ApJ...573...37J,2004MNRAS.350.1322F}.

However most stellar population models do not include the modelling of 
emission lines in galaxies. This is mainly due to the need for modelling
the different phases present in the interstellar medium. 
Some spectrophotometric codes such 
as Pegase \citep{2004A&A...425..881L} do include simple prescriptions
for emission lines. 
Other models such as Starburst99 \citep{1999ApJS..123....3L} actually
provide equivalent width and flux estimates for the most important
emission lines usually observed in galaxy spectra.

The broad band shape of the SED distribution can be related 
empirically to the presence of an emission line. Roughly speaking, we know that 
the emission line strength depends on the condition of the gas inside
a galaxy; this is strongly related to the amount of star formation ongoing
in this galaxy; the star formation can be inferred to a certain extent
from the colour, i.e.  
we know that the SED of a star burst galaxy is much flatter than the 
corresponding SED of a more passively evolving red galaxy. Therefore 
the shape of the SED must be correlated in some way to the presence of 
a strong emission feature in the same galaxy. 

We propose to find empirical correlations between the equivalent
widths of different lines and the broad band colours from imaging data
by utilising novel statistical methods.  One method, Artificial Neural
Networks, is based on the ANNz code
\citep[e.g.][]{2004PASP..116..345C,2007ApJ...666..747B,2007arXiv0705.1437A},
previously used to predict photometric redshifts from galaxy colours.
The second statistical method we explore here is Locally Weighted
Regression (LWR).  We can then use such correlations to understand
better the process of galaxy formation and the conditions under which
strong lines are produced.  We suggest that with these statistical
results, any model attempting to predict emission features in galaxy
spectra could be considerably improved by being calibrated to have the
same statistical properties as found in nature.

Another strong motivation for this work is the predictability of an 
emission feature present in high redshift galaxies. 
The Sloan Digital Sky Survey 
(SDSS) and the 2dF Galaxy Redshift Survey (2dFGRS) have probed the nearby
Universe. More current (VVDS and DEEP2) as well 
as future efforts will attempt to probe a large part of the high 
redshift Universe. More specifically if one is only interested in 
redshift surveys and not on the more specific SED of a galaxy then it 
is important to minimize the amount of time spent looking for a galaxy 
redshift. The most efficient way of doing this is to perform redshift 
surveys targeted at emission line objects. If there is a possibility 
of predicting statistically which galaxies have strongest emission lines 
then the potential time spent locating all these galaxies in redshift 
space can be reduced and the scope for larger surveys 
increased in the same way.
These approaches would be useful in particular for 
the forthcoming FMOS survey \citep{2006SPIE.6269E.136D}
and the planned WFMOS survey.

The outline of this paper is as follows. In Sec.\ref{sec:data} we
present the data used to provide correlations found, we examine a low
redshift sample (SDSS) as well as a high redshift sample (DEEP2).
Section 3 presents the methodology used to extract the correlations
from the data and describes the results found, we attempt to use LWR
and ANNz for this purpose.  
In Section 4 we present methodology for classification of 
AGNs and starforming galaxies.
We comment on the prospects that this
technique has in speeding up future redshift surveys in Section 5 and
conclude in Section 6.  We also provide two short appendices
describing the mathematics of the LWR and ANNz methods used.

\section{The data}
\label{sec:data}

\begin{figure*}
\begin{center}
\includegraphics[width=14.5cm,angle=0]{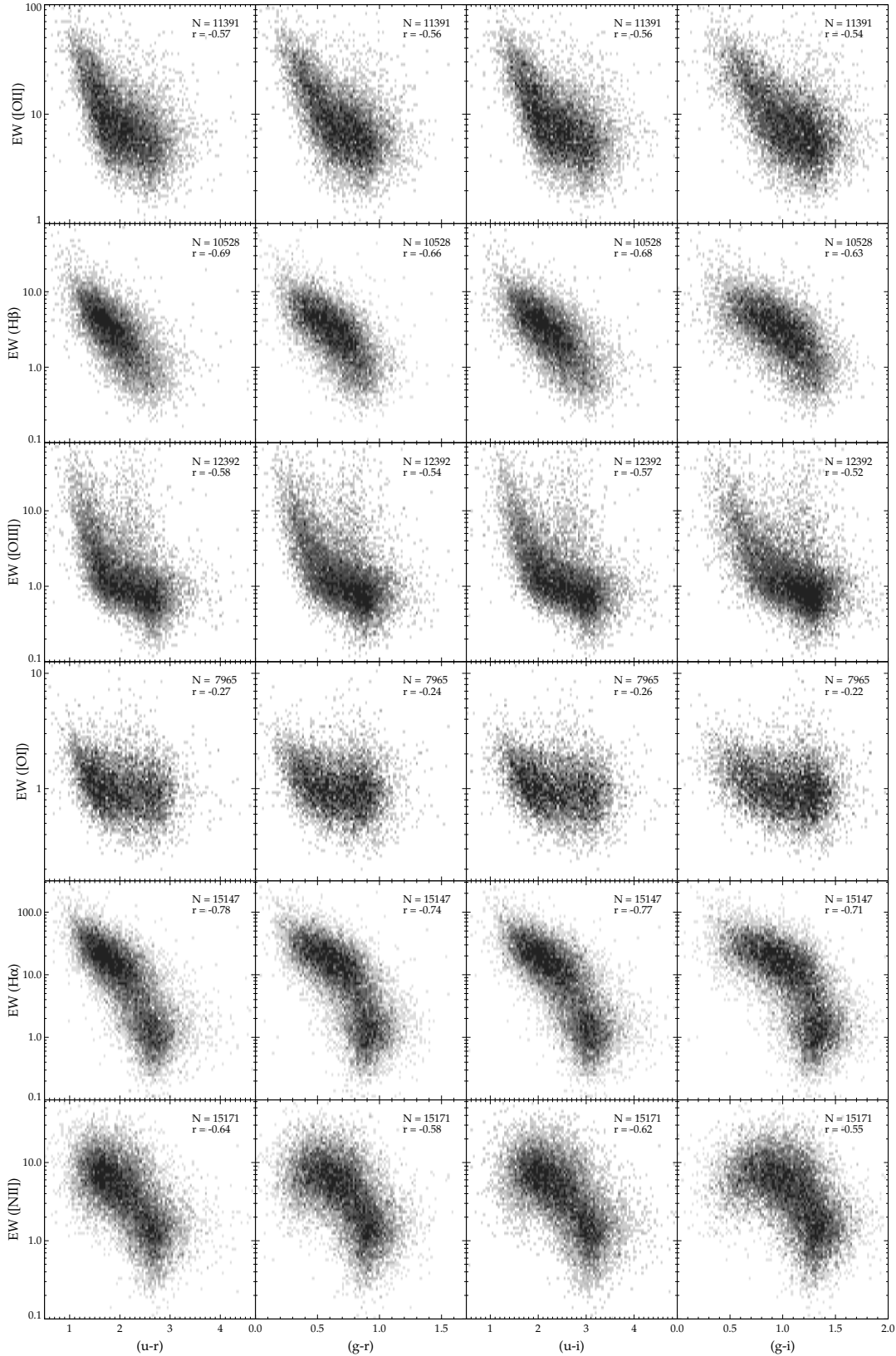}
\caption{Equivalent widths of the emission lines \oii, \Hb, \oiii,
\oi, \Ha, and \nii, measured from the SDSS spectra as a function of
the colours (u-r), (g-r), (u-i), and (g-i). The top-right numbers are
the total number of galaxies in each panel and the Spearman rank
correlation coefficient. The grey scale level represents the number of
galaxies in each pixel, darker pixels having more galaxies.
\label{fig:colour}}
\end{center}
\end{figure*}

\subsection{SDSS}

The low redshift data used in this work is a small sample taken from the SDSS 
Data Release 2 \citep{2004AJ....128..502A}. Photometry is available for all
galaxies in five optical bands ($u$, $g$, $r$, $i$ and $z$). 
We considered the same flux
limited sample described in \citep{2006MNRAS.tmp..840S}. This is a random sample
of 20000 galaxies with reddening-corrected Petrosian r-band magnitudes $r
\leq 17.77$, and Petrosian r-band half-light surface brightness $\mu_{50} \leq
24.5$~mag~arcsec$^{-2}$ \citep{2002AJ....124.1810S}. As a quality cut, we
selected only the objects that show a signal-to-noise (S/N) ratio 
greater than 5 in the $g$, $r$ and $i$ bands. We plot the
redshift distribution of this sample in Fig.\ref{fig:nz}.

The SDSS spectra used here cover the wavelength range $3800$--$9200$ \AA, and 
have a spectral resolution of $\sim 1800$. They were taken with the standard 3
arcsec diameter fibres in the SDSS spectrograph. The spectra are first corrected
for Galactic extinction using the maps of \citet{1998ApJ...500..525S}
and using the extinction law of \citet{1989ApJ...345..245C}. They are
then brought to the rest frame and resampled from 3400 to 8900 \AA\ in steps of
1 \AA\ with a flux normalization by the median flux in the $4010$--$4060$ \AA\
region. These procedures are necessary to the spectral analysis described in
section \ref{sec:lines}

\subsection{DEEP2}

The data sample used at high redshift was the DEEP2 first data release 
\citep{2003SPIE.4834..161D}. The DEEP project is an ongoing project, producing
spectroscopy for targets at a redshift range $0.75 < z < 1.5$. The surveys uses
the DEIMOS spectrograph \citep{2003SPIE.4841.1657F} on the Keck II telescope and
aims to target around 40000 galaxies over 3 square degrees. The targets are
pre-selected from imaging with B, R and I filters taken with the CFH12k camera
on the Canada-France-Hawaii telescope \citep[e.g.][]{2004ApJ...609..525C}.
Galaxies are imaged in $B$, $R$ and $I$ bands and selected to $R_{AB}< 24.1$.
Furthermore, a colour-cut is applied to pre-select high-redshift objects above a
redshift of $z > 0.7$ and only $\simeq$ 3 per cent of galaxies above this
redshift are rejected by the colour cut. The spectra are taken at moderately
high resolution ($R \sim 5000$) and span the range $6300 < \lambda < 9100$~\AA.
Hence the doublet \oii$\lambda$3727 is found from $0.7 < z < 1.4$. This sample
contains 4681 objects and its redshift distribution is plotted in
Fig.\ref{fig:nz}. 

\subsection{Emission line measurements}
\label{sec:lines}

In order to measure the emission lines from the SDSS and DEEP2 galaxy spectra 
we have used a code to fit them as Gaussian functions, composed of three
parameters: width, offset (with respect to the rest-frame central wavelength)
and flux. In the case of the SDSS sample,
we measured emission lines using the methods of 
described in detail by \citet{2005MNRAS.358..363C} and
\citet{2006MNRAS.370..721M}. 
The following emission
lines have been measured: \oii, H$\beta$, \oiii, \oi, H$\alpha$, and \nii.\footnote {In the entire
paper \oii\ stands for \Oii, \oiii\ for \Oiii, \oi\ for \Oi, and  \nii\ for
\Nii.} Lines from the same ion are assumed to have the same width and offset,
and we consider the following flux ratio constraints:
\oiii$\lambda5007$/\oiii$\lambda4959 = 2.97$ and
\nii$\lambda6584$/\nii$\lambda6548 = 3$. To measure the intensities of emission
lines we have to remove the stellar contribution in the continuum at their
wavelength range, mainly to account for absorption features in the emission line
regions. This is done by computing for each SDSS galaxy a synthetic stellar
spectrum obtained from a linear combination of simple stellar population 
spectra that fits the observed continuum in the whole spectral range (after
removal of the zones of emission lines and bad pixels). Removing this synthetic
spectrum from the observed one leaves us with a pure emission line residual
spectrum from which we can easily measure the emission lines. This method
provides a reliable estimate of the stellar
absorption in the entire spectrum, including the windows where emission lines
are found. This procedure is very important since the regions of some emission
lines (mainly the \Ha\ and \Hb\ Balmer lines) can be contaminated by strong
absorption features which may reduce the equivalent width of the lines. For
lines which have no absorption this should make no difference. We also have
compared results with integrating over the spectrum and comparing it to the
continuum and results do not change much showing that the method is robust.

For the DEEP2 data, only the \oii\ line provided suitable measurements. 
Moreover, we have adopted a distinct approach to measure this emission line.
Instead of fitting the continuum with a synthetic spectrum we have performed a
polynomial fit of two continuum windows ($3653$--$3713$ and $3741$--$3801$ \AA)
around the line. The emission line was then measured from the continuum
subtracted spectrum through the Gaussian fitting procedure.

\section{Methodology: Predicting emission features with broad band photometry}
\label{sec:method}

\begin{figure}
\begin{center}
\includegraphics[width=8.5cm,angle=0]{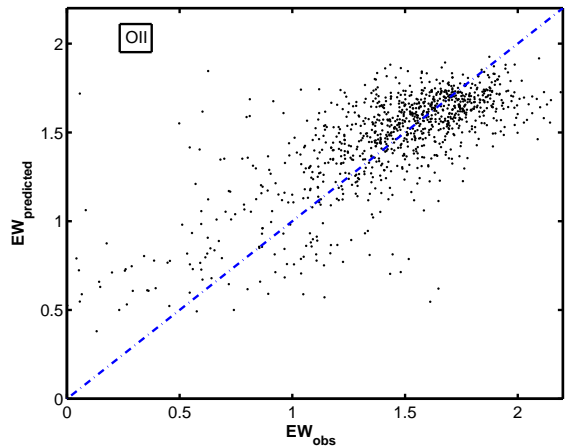}
\caption{The scatter plot for the 1100 galaxies that were used as a testing 
set for the prediction of the equivalent width of the \oii\ line in the DEEP
sample. The Artificial neural network method used in this plot. We can see
a correlation in the log of the predicted equivalent width and the actual
measurement. The scatter in the log is of 0.27. Similar results 
with virtually the same scatter are found with the LWR technique. 
\label{fig::deep _ew}}
\end{center}
\end{figure}

\begin{figure*}
\begin{center}
\includegraphics[width=12.5cm,angle=-90]{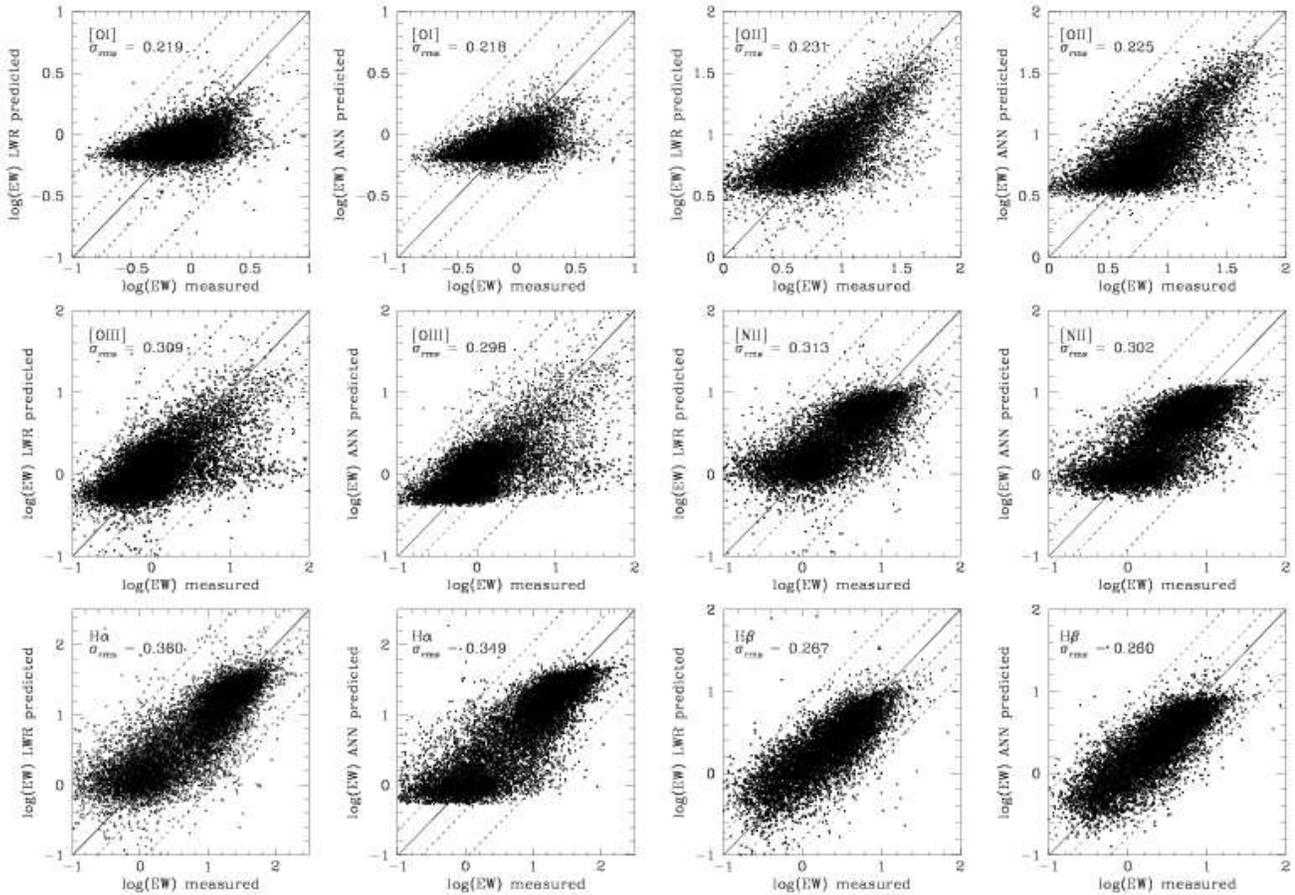}
\caption{The scatter plot of the log of the real equivalent width versus the
predicted equivalent width of galaxies for different lines. For all the plots
we have used only galaxies in the testing set which have not made part on the
training of the machine learning algorithm that has made the prediction.
We present here results using both methods described in this paper. 
As we can see the results found with both results are comparable
and we found no evidence that one method performed better than the other
given the data volume available. We can also see that some lines 
present a better correlation than others and that generally recombination
lines are better predicted than collisional lines. 
\label{fig::scatter}}
\end{center}
\end{figure*}

It is well know that galaxy colours present a good correlation with
the emission line properties of galaxies, such as their equivalent
widths (EW). We illustrate this in Fig. 2, where the equivalent widths
of emission lines measured from the SDSS spectra are plotted as a
function of various galaxy colours. Several correlations can be
identified between these quantities, as confirmed by the high values
of the Spearman rank correlation coefficients obtained for some of
them, particularly those involving the \Ha\ and \Hb emission lines.
Therefore, it is not only the 4000 \AA\ break colour indice that
encodes information about the emission features in galaxies (e.g.
Mateus et al. 2006), as typical galaxy colours can also be used to
infer these properties in a convenient manner, from photometric data
only. On this basis, here we use different methods to combine all
information available in the galaxy colours to produce an empirical
relation between colours and emission strength.

We have analyzed the data with two different techniques to reinforce
the robustness of the correlations found and to compare the two
techniques used.  We have used Artificial Neural Networks
\citep{Bishop}, as implemented in the publically available ANNz code
(Collister \& Lahav 2004 and refereneces therein) and a Locally
Weighted Regression (LWR) \citep{LWR}.  See the Appendices for the
mathematical details and implementation of these methods.

Both of the techniques we present here rely on a training set which 
is representative of the true population of galaxies in order to retrieve 
the information on the line features in each galaxy. We have separated
the sample into two groups, one group which we have used to train the 
machine learning algorithms and a second sample which we will name the 
testing set which is set aside and then used to test the reliability of the 
methods. We have set aside 1100 galaxies in DEEP and 
14000 galaxies in the SDSS as testing sets.

The rest of the sample was used as to train the algorithms, that is 
2000 galaxies to train in the DEEP sample and 6000 galaxies to train in the 
SDSS sample. For both methods used here we have to subdivide this 
sample into a training sample and a validation sample. In the case of
neural networks this is done 
to prevent over-fitting. In the case of the LWR method, this is done to
obtain the best kernel value K which is appropriate to the data. 
Both prescriptions are explained in the Appendix. The architecture
of the neural network used is described in the Appendix and is
of the type N:2N:2N:1 where N is the number of magnitudes available 
which is three in the case of DEEP data and five in the case of SDSS. 

With the DEEP sample the only line which we have produced a fit for was the 
\oii\ line. We have performed this analysis only on this line because the 
DEEP survey relies heavily on the \oii\ line for redshift estimation, 
therefore having a high completeness. Other lines do not appear in a 
large fraction of the galaxies in the entire sample. 
In the sample from SDSS we performed this analysis for all measured lines: 
\oii, H$\beta$, \oiii, \oi, H$\alpha$, and \nii.

We plot in Fig.\ref{fig::deep _ew} our attempts to recover the 
equivalent width of the \oii\ line in the DEEP survey using broad 
band photometry only. We have plotted a scatter plot of the real 
equivalent width versus the predicted equivalent width for the testing set
given the training set.
As we can see the logarithm of the equivalent width is
predicted reasonably well and show that this weak correlation
can be inferred very well with the techniques we have proposed here.
We can predict the log of the equivalent width with an error of 0.27 without 
any spectroscopic information on the redshift of each galaxy.

We present in Fig.\ref{fig::scatter} the results we find by fitting 
different line widths to SDSS galaxies. Again in these plots we have used
only galaxies in the testing set which have not been used in the training
process. We find that some line equivalent widths are relatively well 
predicted by galaxy magnitudes and colours. This is the case for the hydrogen
lines where a reasonably strong correlation was found. 
We have found however that the collisional lines observed are 
harder to predict. For instance the \oi\ line equivalent width had virtually 
no correlation with any combination of the colours and magnitudes found.

We compare results for both of the techniques 
using the same testing sets in Fig.\ref{fig::scatter}. 
We plot them side by side so that we can see
the relative comparison between the neural network non-linear fit to the data 
and the locally linear regression used. We have inspected the data and found that
there was no significant difference between both methods. Both predictions 
have yielded results with a very similar scatter given the data 
volume available. We conclude that for fitting purposes neural networks and
a Locally Weighted Regression performs well and on an equal footing.

We have also attempted to simply classify objects into line-emitting and non 
line-emitting. We have made this separation for each galaxy upon 
the simple criterium: whether an emission line is detected above a certain
value. This separation has been done for each line emission 
analyzed. For instance if a certain galaxy has an \oi\ line but does not posses
an H$\alpha$ line then it is considered as an emitting object for the purpose
of the analysis on the \oi\ line but it is considered as a non-emitting 
object for the purposes of the H$\alpha$ line analysis.

\begin{figure}
\begin{center}
\centerline{\includegraphics[width=4.5cm,angle=0]{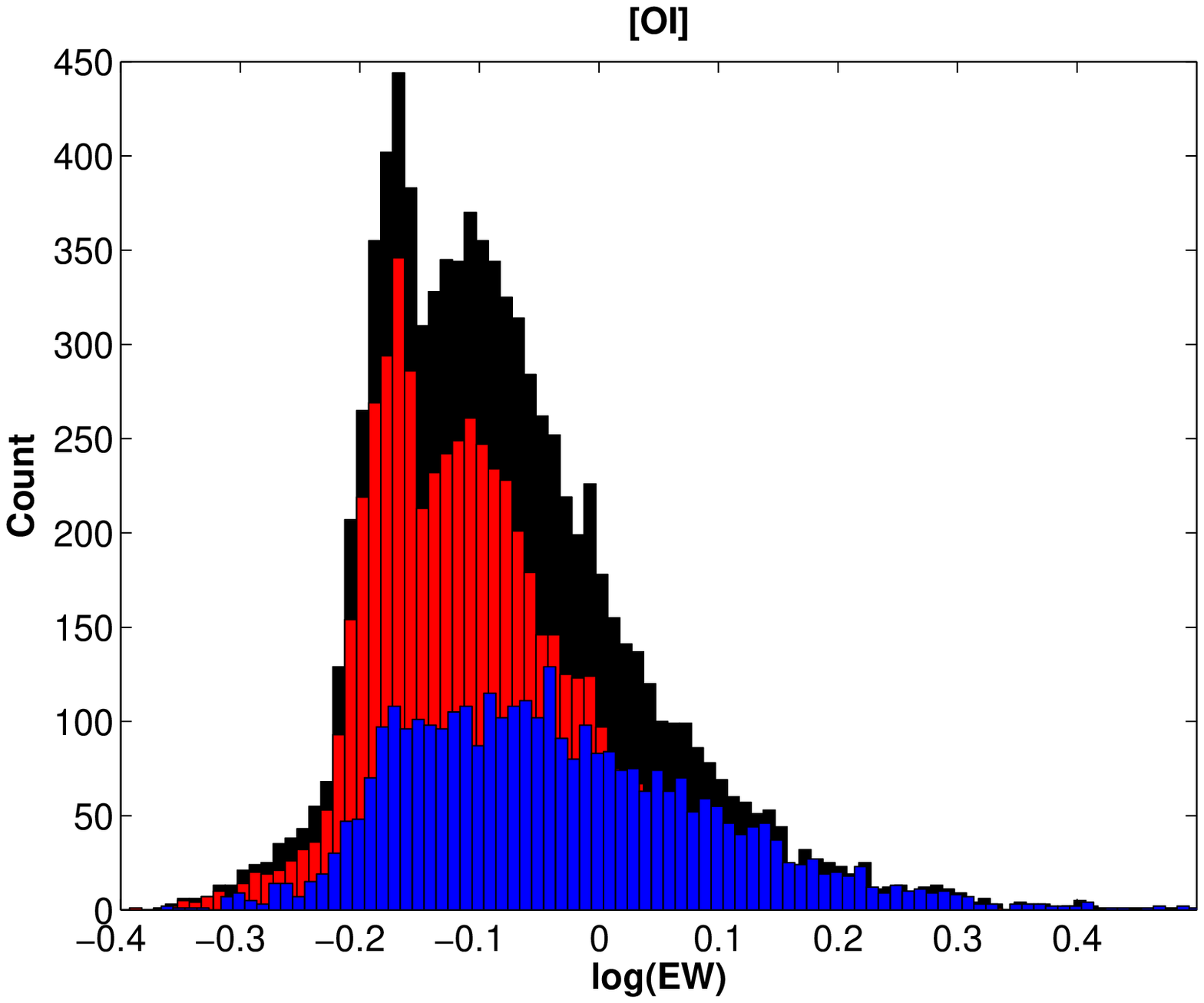}
\includegraphics[width=4.5cm,angle=0]{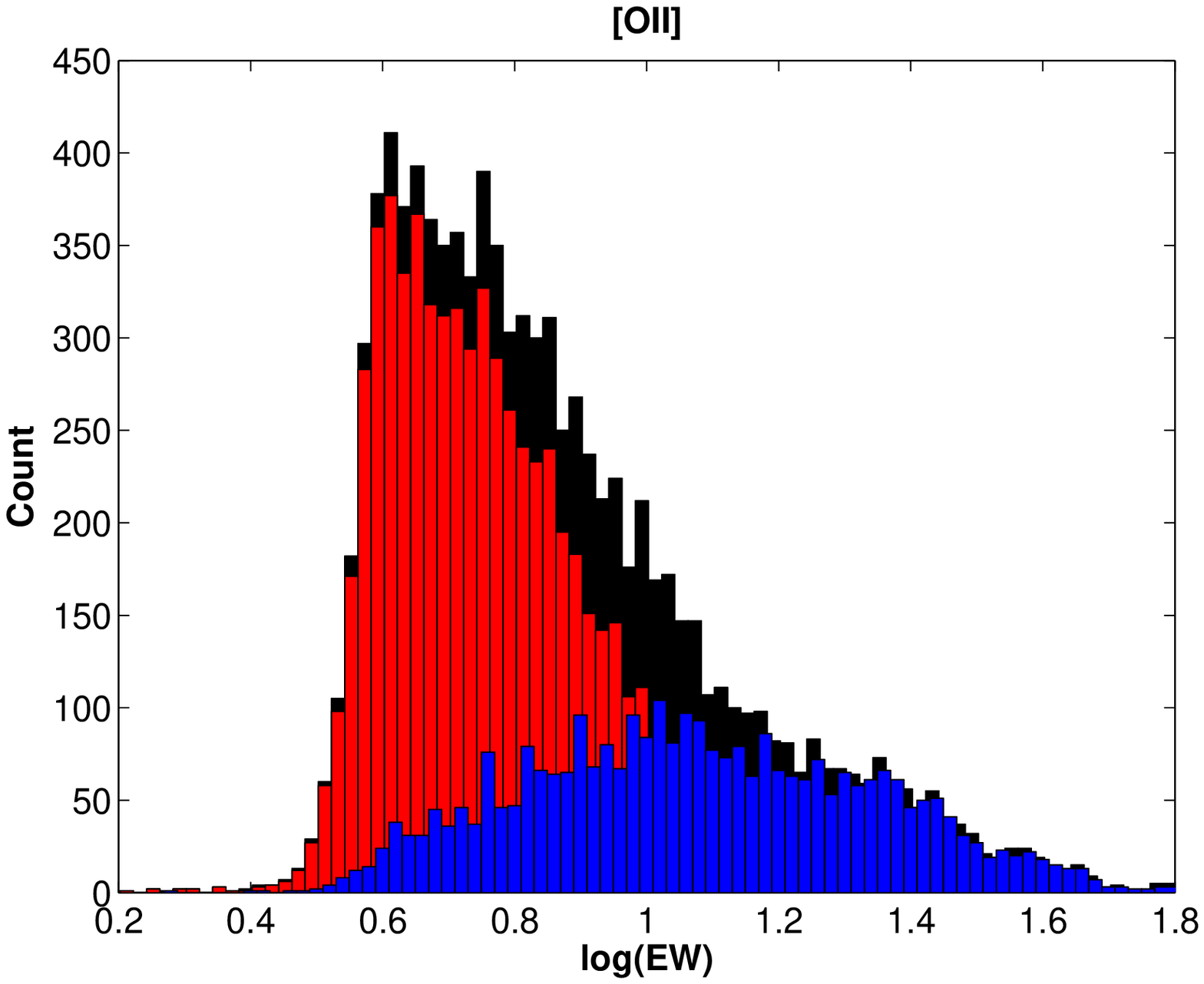}}
\centerline{\includegraphics[width=4.5cm,angle=0]{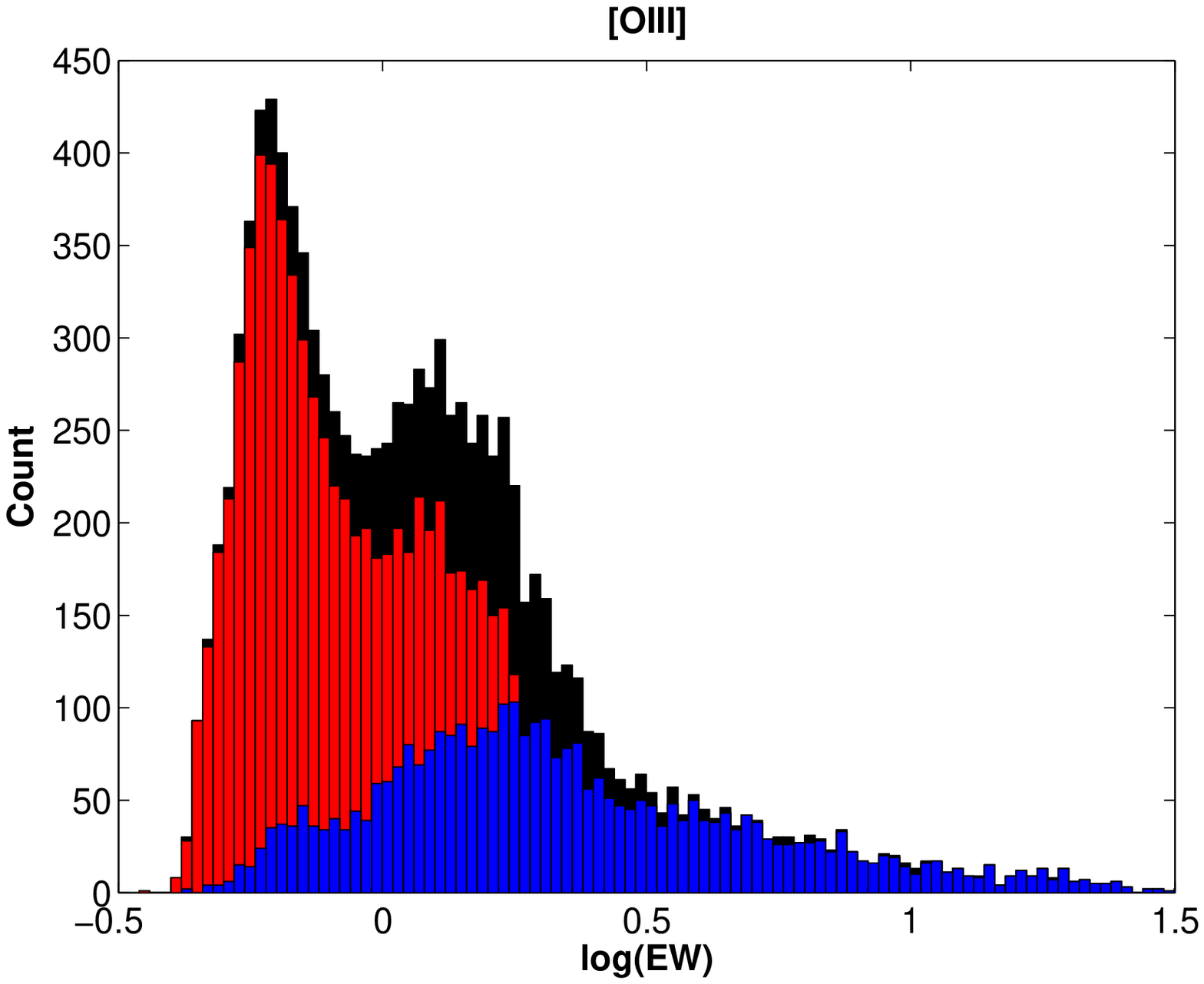}
\includegraphics[width=4.5cm,angle=0]{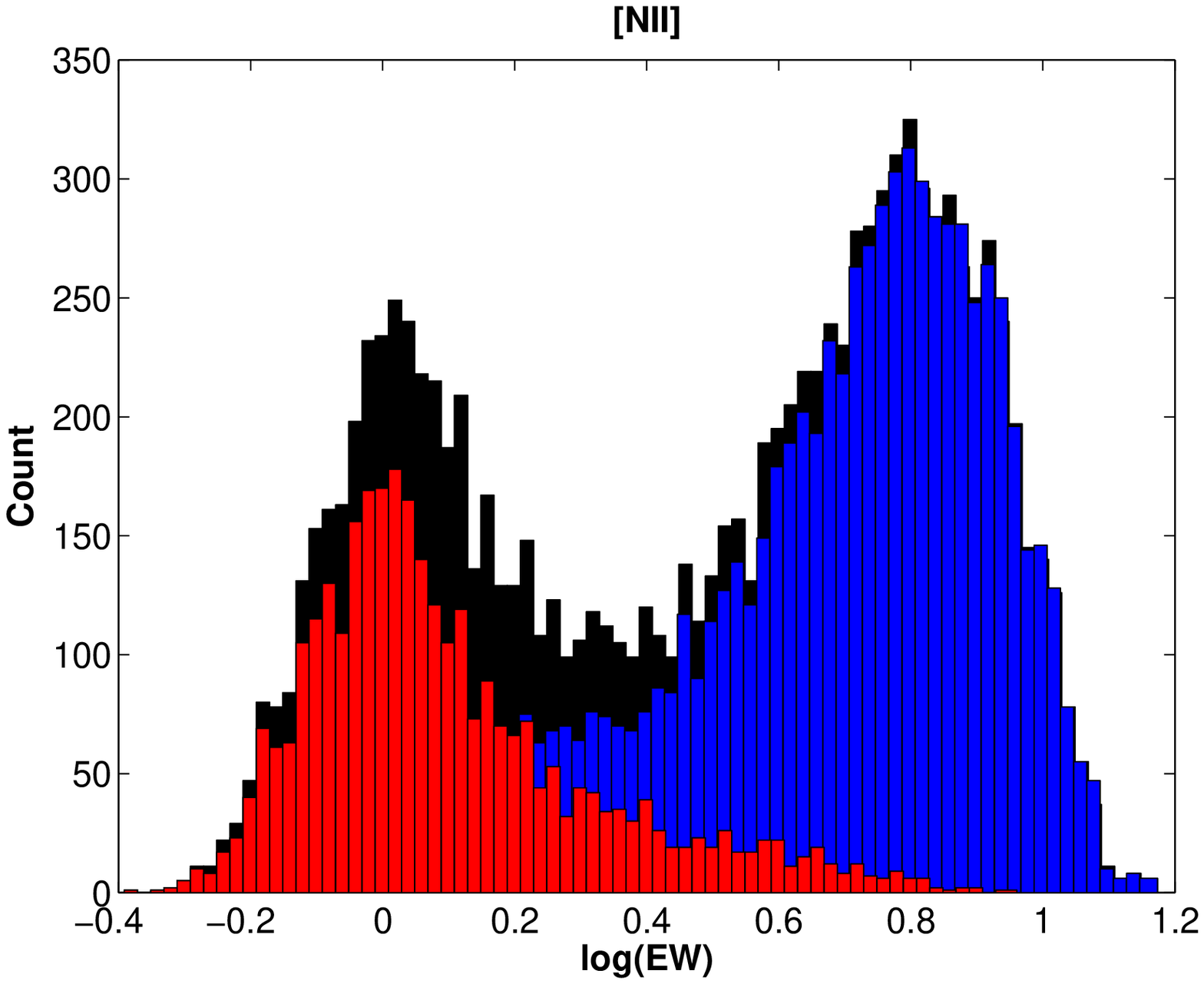}}
\centerline{\includegraphics[width=4.5cm,angle=0]{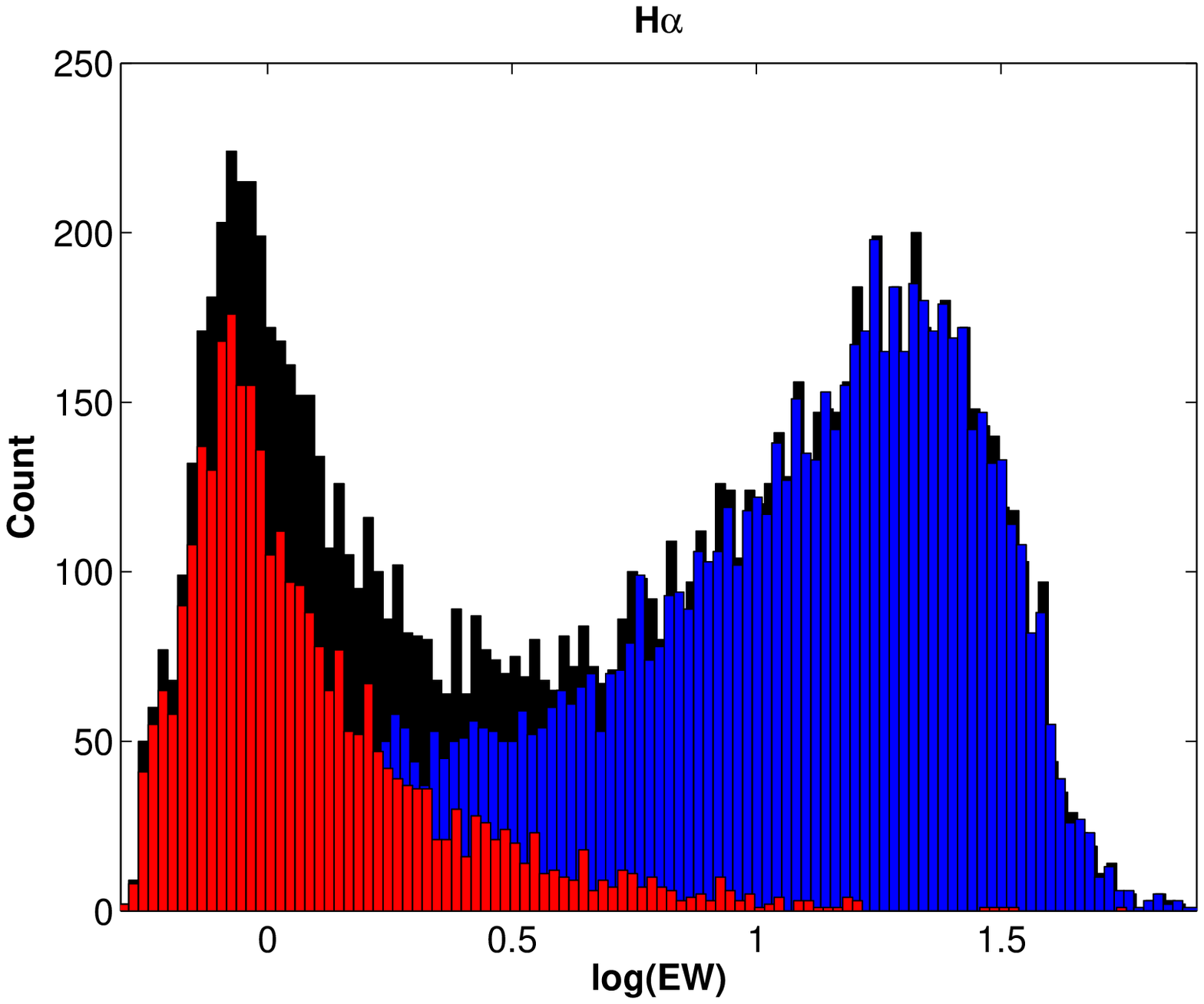}
\includegraphics[width=4.5cm,angle=0]{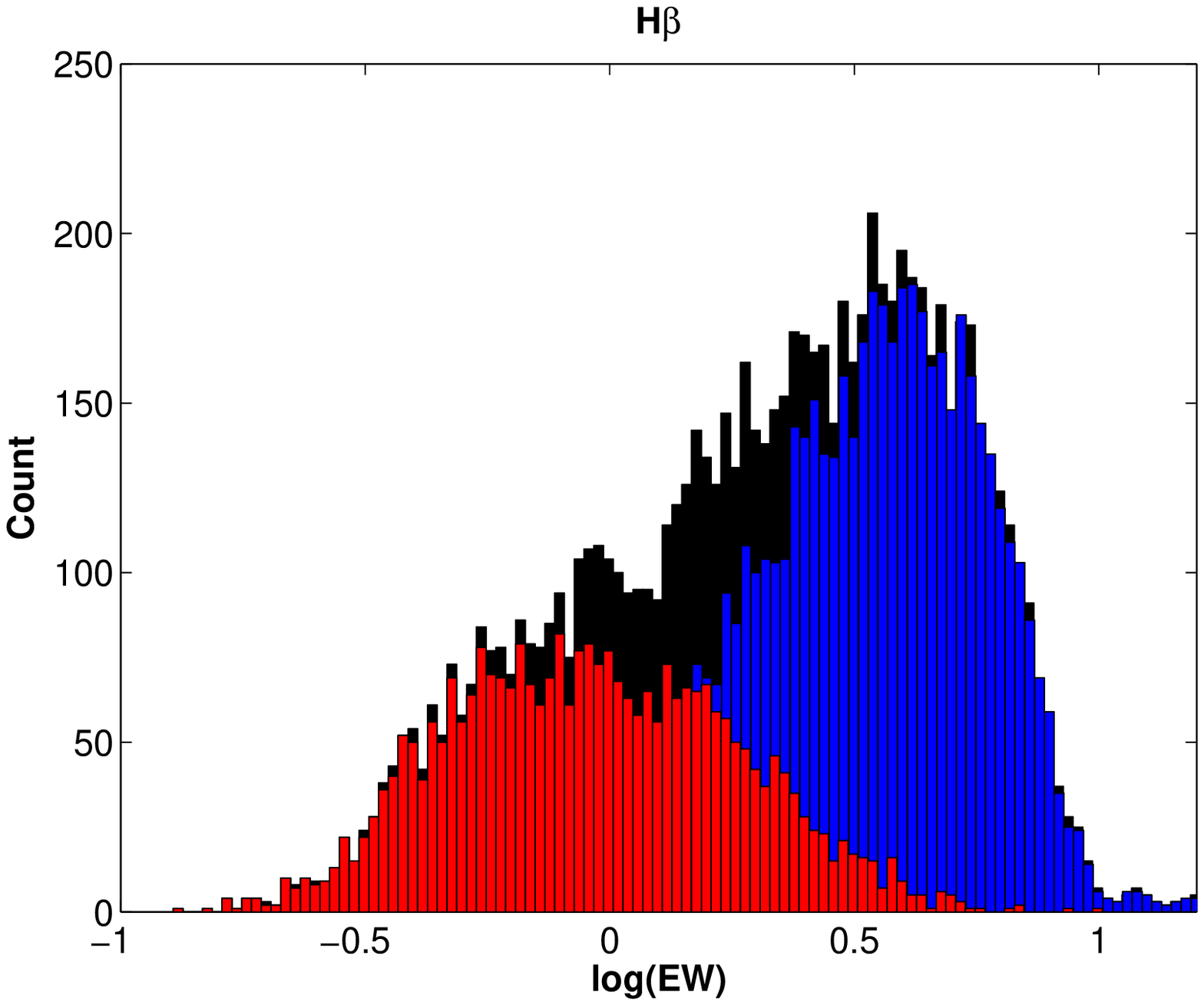}}
\caption{Histograms of the log of the equivalent width for SDSS galaxies.
Each plot represents the results for one line. The black histogram
are for all the objects for which a line was detected in the analysis
of the SDSS data. we have then chosen a median value for the log of
the equivalent width for these lines and separated the black samples into 
a sample with predicted equivalent width higher than this median value
and a sample with the predicted value lower than this median value.
This shows the amount of overlap that there is in predicting 
line features with the methods proposed. We can clearly see bi-modality
of galaxies and by looking at the H$\beta$ results we can also
see that our methods go further than simply 
detecting this bi-modality clearly separating objects with high 
and low equivalent width.
\label{fig::hist}}
\end{center}
\end{figure}

\begin{figure}
\begin{center}
\centerline{\includegraphics[width=4.5cm,angle=0]{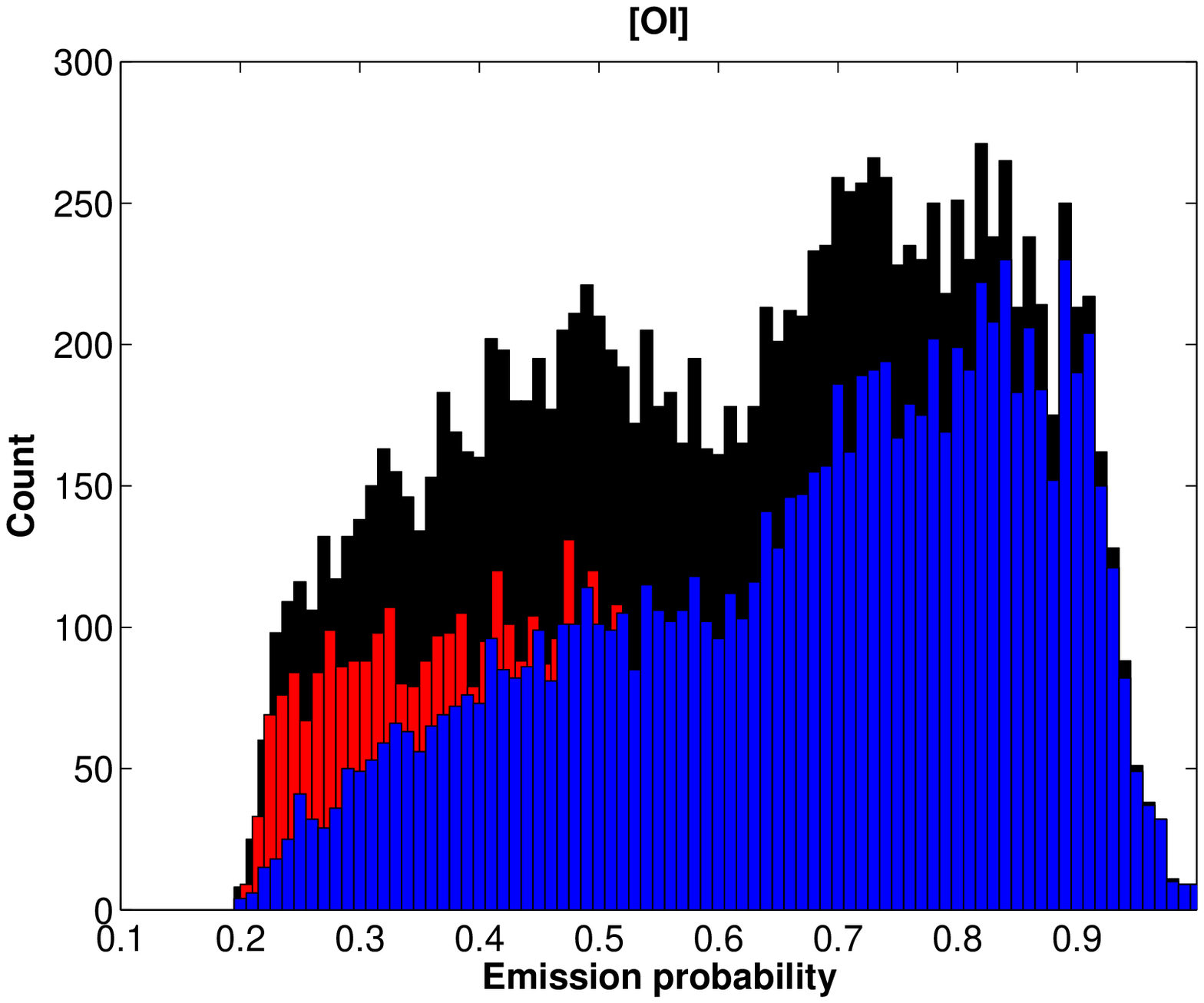}
\includegraphics[width=4.5cm,angle=0]{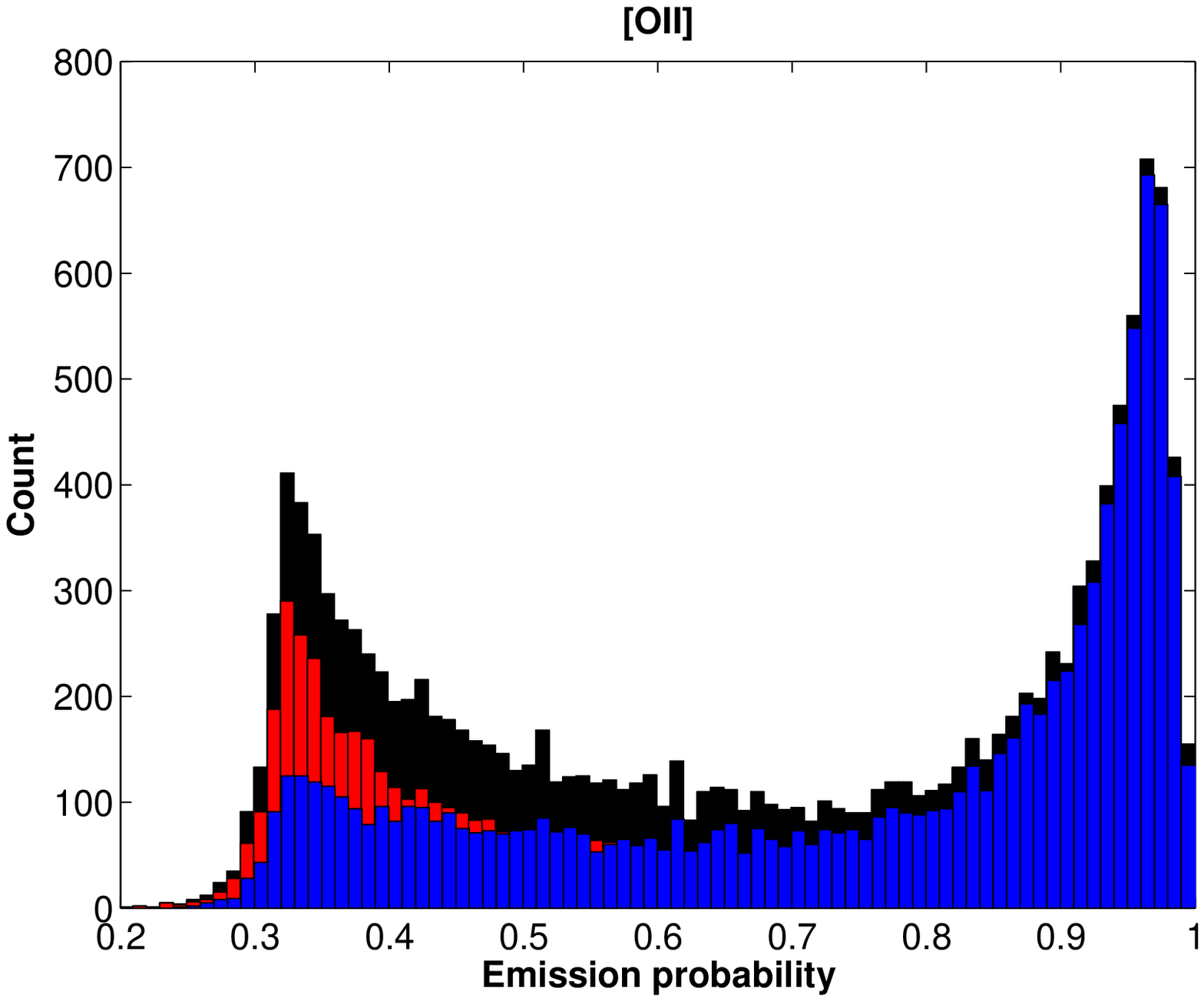}}
\centerline{\includegraphics[width=4.5cm,angle=0]{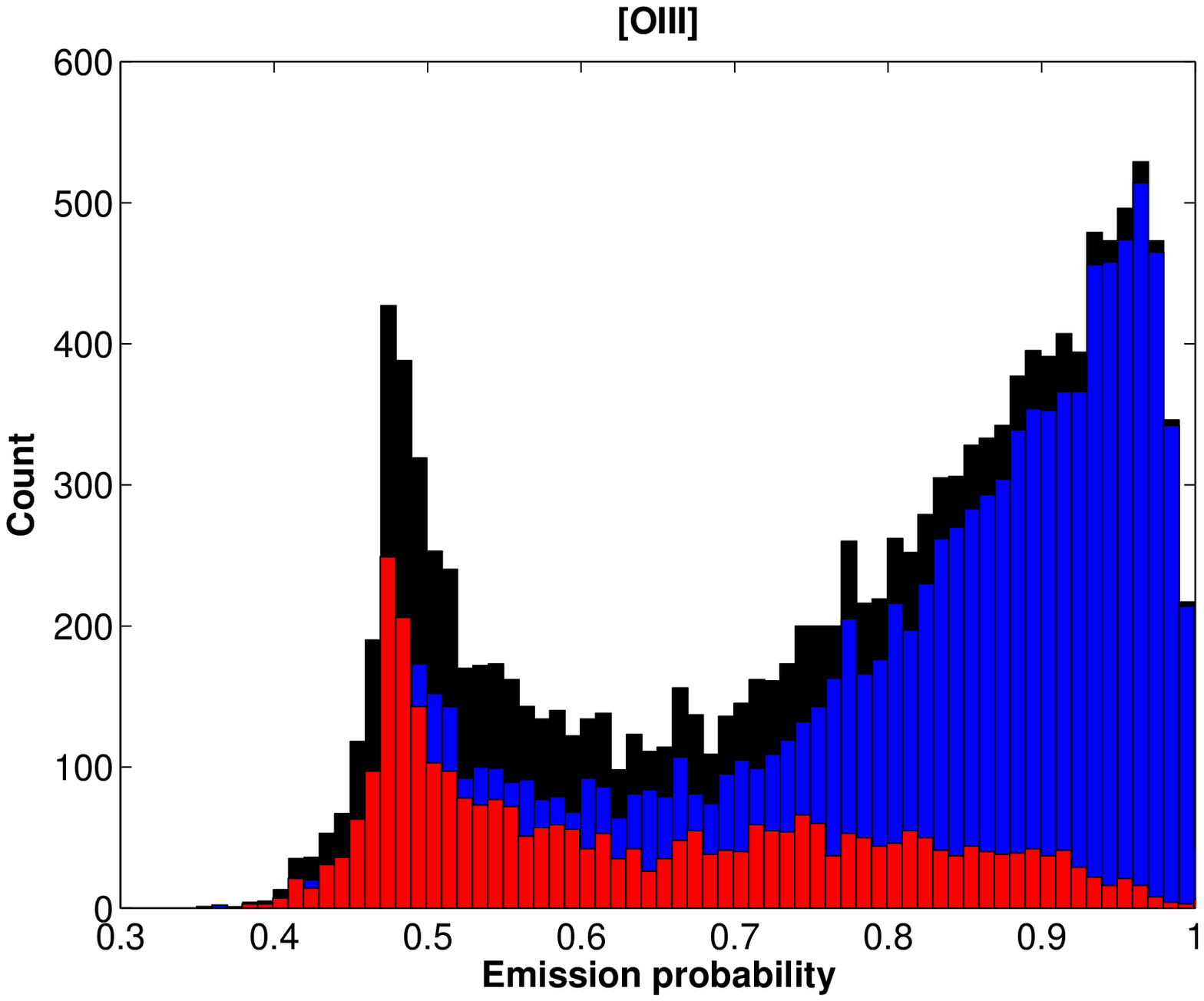}
\includegraphics[width=4.5cm,angle=0]{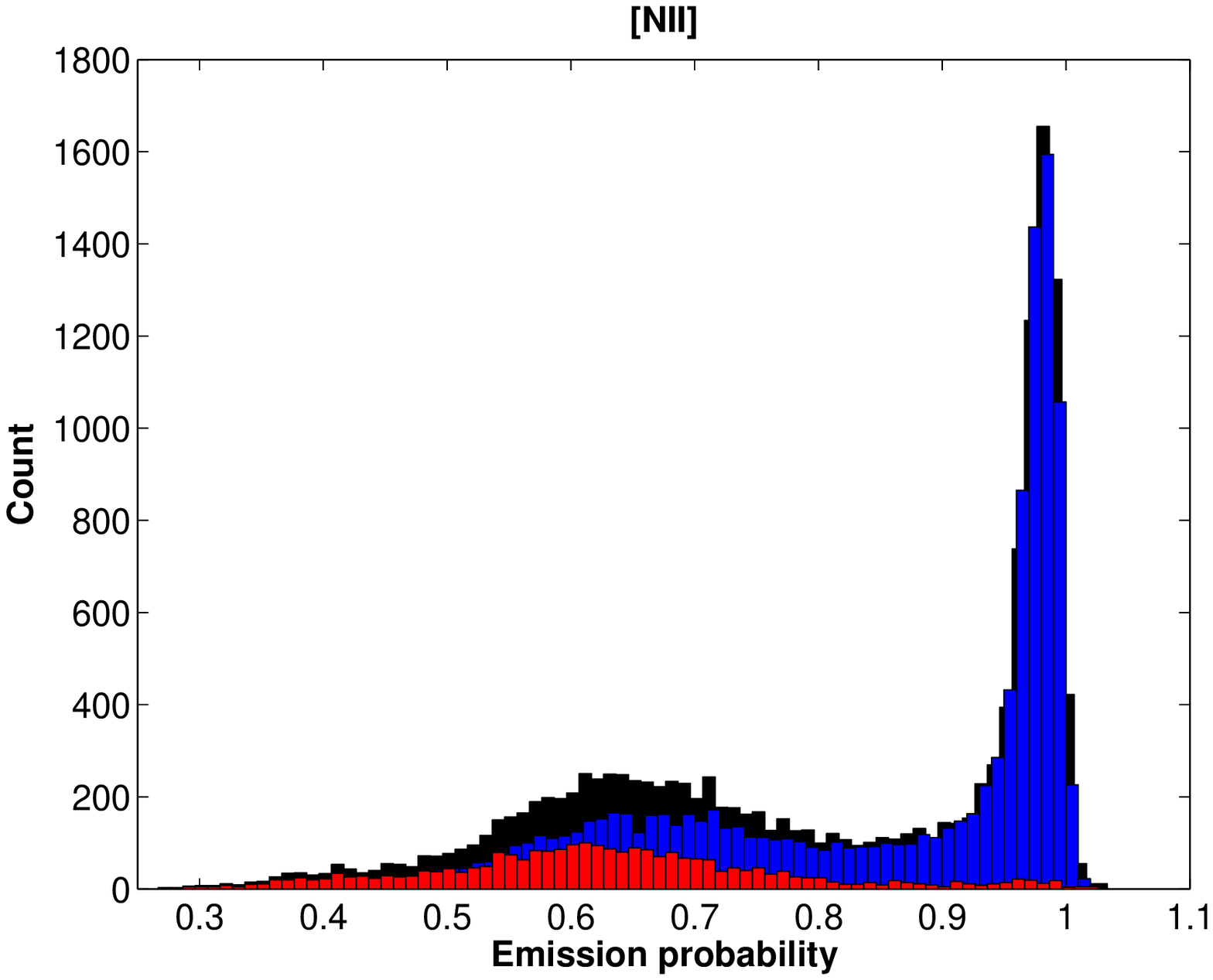}}
\centerline{\includegraphics[width=4.5cm,angle=0]{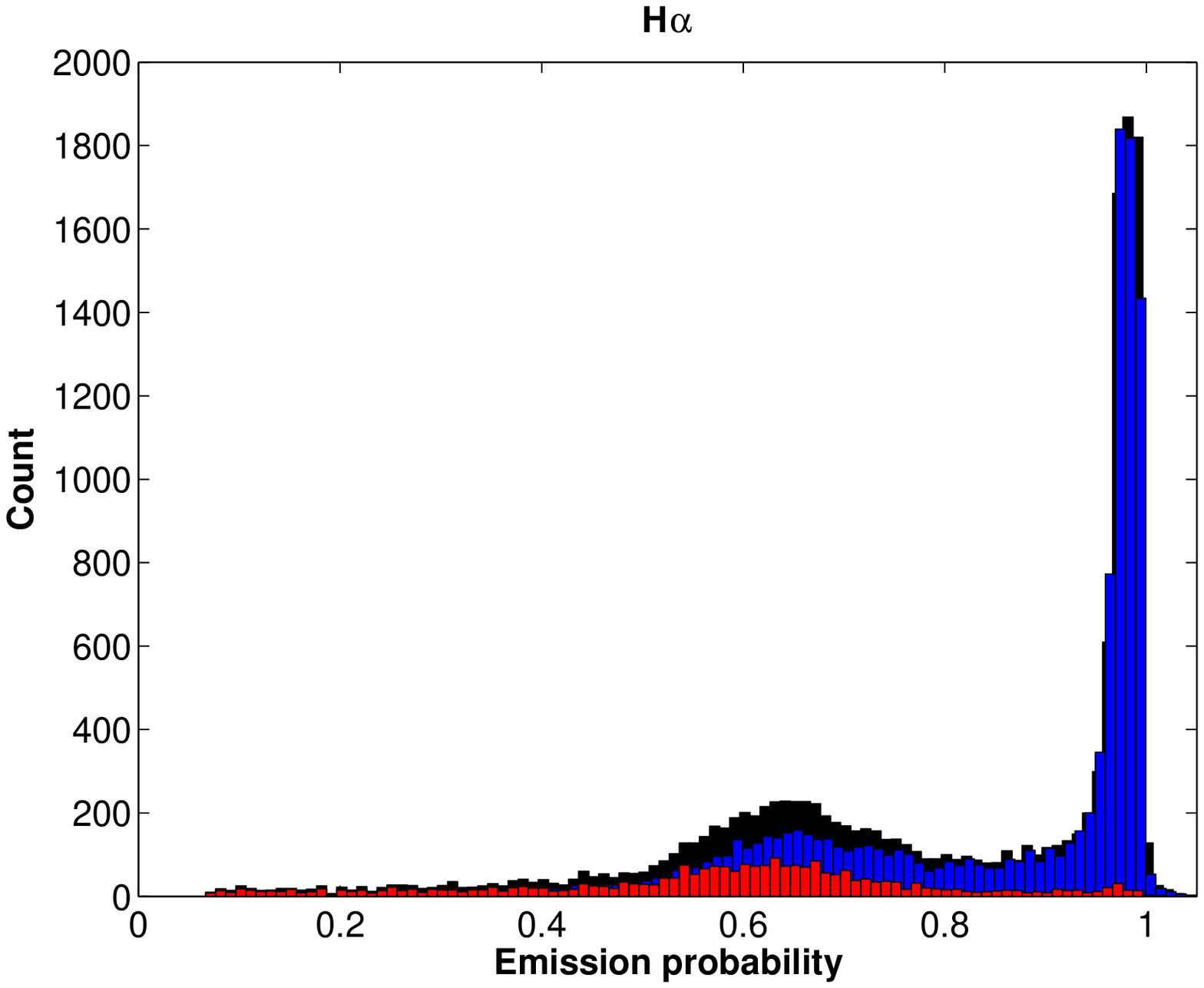}
\includegraphics[width=4.5cm,angle=0]{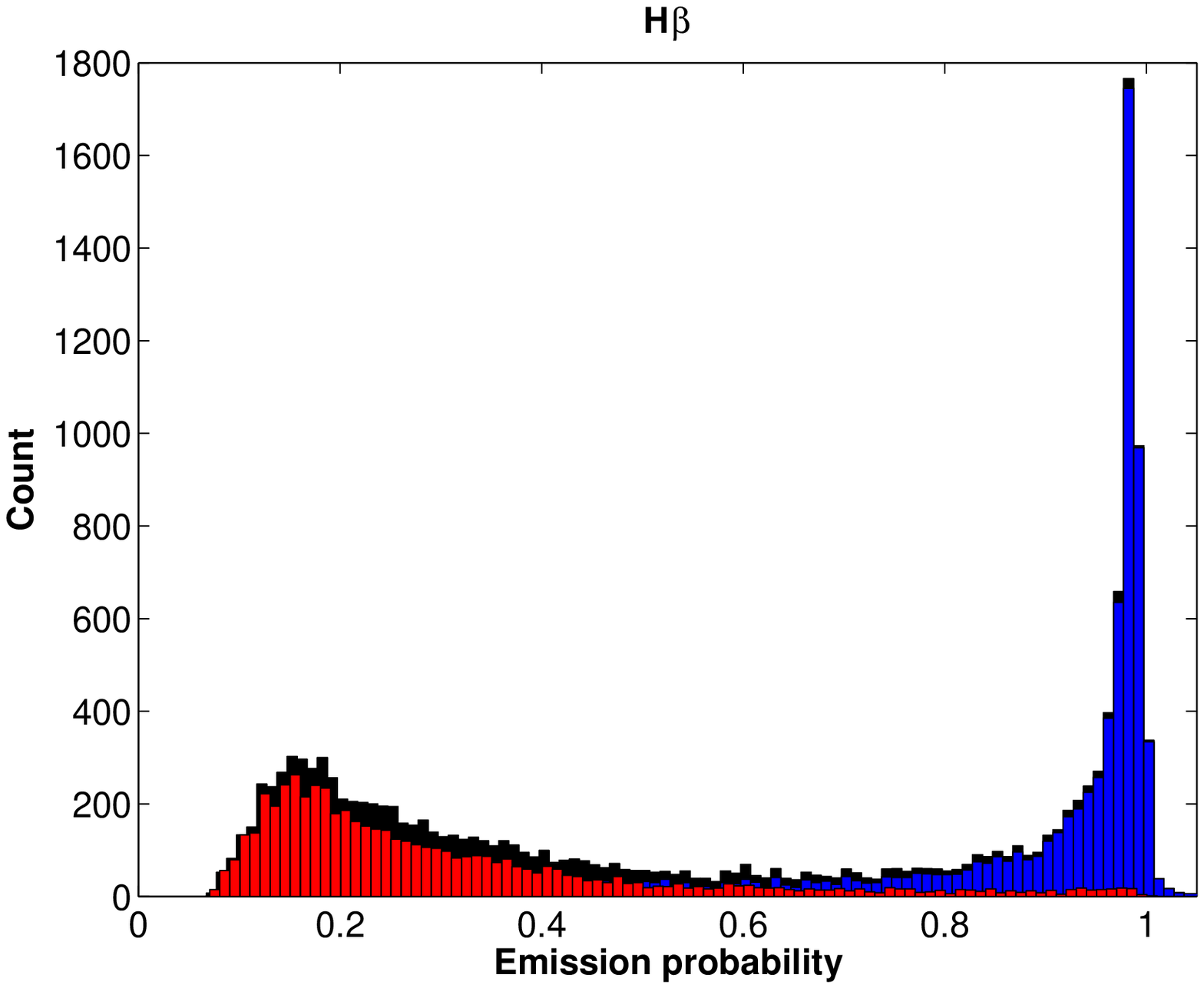}}
\caption{Histogram of the probability that a galaxy is a line-emitting galaxy 
or a non line-emitting galaxy given by the neural networks. The black histogram
represent the entire population. The red and blue histograms represent the line
emitting objects and the non line emitting objects. We can see that we can 
clearly separate the blue and red points for some of the cases whereas other 
lines cannot be diagnosed with this method. 
\label{fig::counter}}
\end{center}
\end{figure}

We have then used the training set in the following way: we have assigned 
the value of one to emitting objects and the value of zero to 
'non-emitting' objects, defined by the line being detected in our
code above a signal to noise of three. 
A neural network was then trained in this specific 
training set. When a neural network is trained by binary values between zero 
and one, it 
returns a probability (e.g. Lahav et al. 1996 and refereces therin) 
of that given object belonging to the class 
described by the binary number. So in our case the value returned by 
the neural network will be the probability of that object having the emission 
feature tested for. We have also attempted to perform the same analysis 
with the LWR technique but found that despite the LWR technique producing 
predictive results which are comparable and as good as the neural network 
fit it was not able to perform this classification task nearly as well as 
the neural networks. This was to be expected as the LWR method is not 
directly applicable to classification tasks. This is because the linear 
fit results are not representative of a probability range in the same 
way the neural networks are.

We plot these results in Fig.\ref{fig::counter}. We can see that the 
results are varied and depend on the emission line being assessed. For 
instance there is a good prospect of training a neural network to distinguish 
between Hydrogen emitters and \nii\ emitters at low redshift but on the 
other hand the \oi\ line exhibits a very poor correlation between colours 
and magnitudes and its EW. The other oxygen lines exhibit some 
correlation that are picked up by our methods.

In order to have an idea of what contamination would be introduced but 
performing a selection based on the predicted equivalent width of a galaxy
with our method, we have plotted histograms for the entire sample
and then chosen the center of the distribution of the equivalent widths 
in log space. We have then plotted the histograms for the samples with 
high predicted equivalent width and low predicted equivalent width in 
Fig.\ref{fig::hist}. 

While predicting the equivalent widths for these emission lines we have 
found the already known result
of galaxy bi-modality. We know that galaxies are bimodal and 
that late type galaxies have higher star 
formation rates and therefore stronger emission features whereas
the opposite is true for late type galaxies. We can see this 
bi-modality in the histogram of the equivalent widths for the different
lines in Fig.\ref{fig::hist}. This is not found on every set of lines because
there are selection effects (fainter lines are simply not seen 
in some galaxies).
For instance the ratio of the H$\alpha$ and H$\beta$ lines 
for a single galaxy should be roughly 
constant just below three. This is because both lines are recombination lines
and the temperature dependence of the emissivity is roughly the same for
both lines. However we do not find this bimodality for H$\beta$ but we find 
it for H$\alpha$. This is simply because the H$\beta$ flux is smaller and this
makes so that a lot of the fainter lines remain undetected.
Furthermore we argue that
the correlations found in H$\beta$ are strong which suggests that 
the method we are using is going beyond simply separating the bi-modality and
is classifying the line widths according to their size.

\section{Methodology: AGN/Star-Formation Classification}
\label{sec:agn}

\subsection{Traditional Spectral galaxy classification}
\label{sec:class}

We adopted a traditional procedure to classify galaxies according to
their emission line properties. By examining diagnostic diagrams formed by line
ratios of optical emission lines,
such as the \oiii/H$\beta$ versus \nii/H$\alpha$ diagram proposed by 
\citet{1981PASP...93....5B}, we can distinguish emission-line galaxies
according to the mechanism responsible for producing the lines. 
In such diagrams,
hosts of active nuclei (AGN) and star-forming galaxies form two very
distinct branches, or wings, making easier the task to separate them. We use
this `convenient' diagram to classify galaxies following the classification
scheme discussed in \citet{2006MNRAS.tmp..840S} based on a theoretical curve
used to distinguish galaxies with pure star-formation from those objects with
some contribution from nuclear activity to the line intensities. Galaxies below
such a curve are classified as normal 
star-forming galaxies (SF). In addition, we
have also used the empirical curve proposed by \citet{2003MNRAS.346.1055K} to
identify the AGN hosts. Galaxies between the two curves are hybrid objects
(the contribution of the AGN to the \Hb emission of galaxies below the Kauffmann
et al. line is at most 3 per cent) and therefore will not be included in our
analysis. The AGN hosts are then selected as those objects located above the
Kauffmann et al. empirical curve.

\begin{figure}
\begin{center}
\centerline{\includegraphics[width=9.5cm,angle=0]{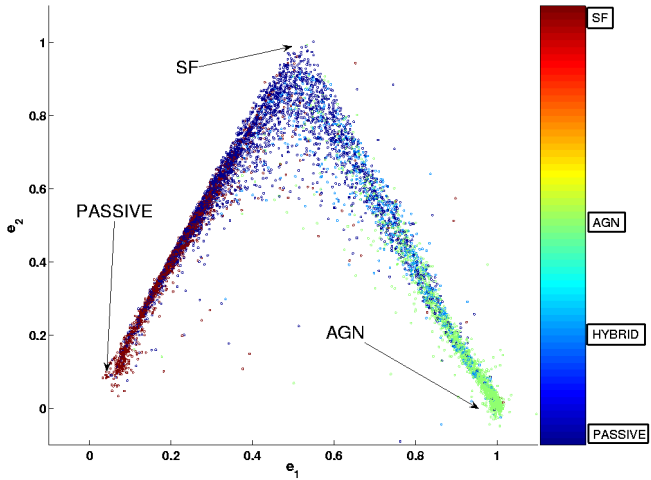}}
\caption{Classification of objects as AGN, SF or passive objects via 
neural networks. the points are colour coded according to the spectral 
classification blue, green, orange, red noting passive galaxies, 
AGN, hybrid objects and star forming galaxies respectively.
Each classification has been assigned a doublet in the following way:
passive galaxies (0,0), AGN (1,0), SF galaxies (0,1) and hybrid 
objects (0.5,0.5). The neural networks has been used to estimate this 
and each point is a galaxy is plotted at
$(e1,e2)=(e1+e2/2)\,{\bf i}+ e2\,{\bf j}$. As we can see it is possible to
predict AGN features from broad band photometry well. There is a degeneracy 
between the colours of many passive galaxies and star 
forming ones but it is in general possible to classify them 
with colours only. As expected hybrid objects lie on the 
line between AGN and SF galaxies.
\label{fig::agn}}
\end{center}
\end{figure}

\subsection{Automated Spectral galaxy classification}
\label{sec:class_auto}

In this subsection we attempt to classify the spectral features instead 
of simply predicting the equivalent width of lines. We have
described in Sec.\ref{sec:class} how we separate galaxies with emission
features into AGN or star forming (SF) galaxies. 

We have separated the 20000 SDSS galaxies we used in this paper 
into the following classes: passive galaxies, AGN, star forming galaxies
and hybrid objects. Some objects were not classified by the code analysing the
spectroscopic data and we have removed those galaxies. We have associated
a doublet $(e1,e2)$ to each one of these categories in the following way:
passive galaxies (0,0), AGN (1,0), SF galaxies (0,1) 
and hybrid objects (0.5,0.5). 

We have then used a neural network with 5 input nodes, 2 hidden layers with
10 nodes in each layer and 2 output nodes to classify galaxies into
AGN, passive galaxies and SF galaxies according to the doublets assigned above.
We have used only photometry to perform the training. We used 5000 galaxies 
for training and 3000 for validating the network. The remaining galaxies were 
plotted as a testing set in Fig.\ref{fig::agn}. The network returned a doublet
for each galaxy. For visual reasons, we plot the doublets in the following
basis $(e1,e2)=(e1+e2/2)\,{\bf i}+ e2\,{\bf j}$, so that the labels 
AGN, SF galaxies and passive galaxies are on the corners of an equilateral
triangle.

As we can see from Fig.\ref{fig::agn}, there is a trend on the distribution of
points in this $(e1,e2)$ diagram. The distribution does 
not infer a transition between Passive galaxies to SF galaxies to AGN. 
It indicates that the colour data is able to make a good distinction
between passive galaxies and AGN, there is a relativelly good distinction
between AGN and SF galaxies and a certain 
overlap between passive and SF galaxies, which we had already found 
attempting to classify galaxies in the previous section. This offers 
an alternative to classifying objects photometrically as AGN or star forming
galaxies via a training set and broadband photometry only.

\section{Prospects for future surveys.}
\label{sec:discuss}

We have shown that it is possible to predict emission features using broad band
photometry. We find that the correlations vary in strength from line to line. 
Particularly recombination 
lines are more predictable than collisional lines. We have not included any
information on the shape or size of the galaxies. We expect that the results 
found here would be stronger if information such as shapelet coefficients for
galaxies is included into the analysis as we know that galaxies with different
star formation rates and therefore different line strengths have different
morphologies.

We argue that this technique can be used in future spectroscopic
redshift surveys to speed them up. Future facilities such as FMOS or WFMOS
will produce spectroscopic surveys in the optical and the IR part of the 
spectrum targeting mainly the \oii\ and the H$\alpha$ lines. We have shown here 
that there is a reasonable predictability of the \oii\ line at high redshift 
and that the H$\alpha$ and H$\beta$ line are well predicted in the low redshift
Universe. A high redshift complete sample of around 10000 objects could be 
used to train a network to then select the preferred targets.

If one chooses to predict the line strengths one alternative would be to fit 
stellar population synthesis models to the observed spectrum and from that
infer a star formation rate and therefore a line flux. We argue that the
technique we have proposed here is complementary and one would be able to 
encode information on the morphology of the galaxy easily whereas if one 
fits stellar population models to the training data it would be hard to 
include special information in the analysis.

\section{Concluding remarks}
\label{sec:conclusion}
 
We have looked, in this paper, for empirical relations between the 
equivalent width for several different lines and the broad band colours
for the same objects. We used an automated way to 
explore the correlations found in the data
via the use of 
LWR and ANN. The two methods give very similar results. 

We have performed the analysis in two samples, a low redshift sample obtained
from the Sloan Digital Sky Survey covering the nearby Universe and
a high redshift sample 
from the first data release of the DEEP survey covering a 
deeper sample off the Universe. We found that in the DEEP data 
one could predict the equivalent width of the \oii\ line
relatively well by looking at broad band colours only. With the SDSS
data six lines were predicted from a training set sample. In general 
collisional lines presented little correlation although some of them
were reasonably predicted by the colours. There was a stronger 
correlation found in recombination lines.

We have compared the power of prediction of both methods used. We have 
concluded that both Artificial Neural Networks and the Locally
Weighted Regression methods are capable of recovering most of the information
encoded in the training set with little statistical difference between
both methods. We have used both methods for classification purposes
and found that the Neural Networks are capable of classifying 
the objects into line-emitting and non line-emitting but that
the Locally Weighted Regression method was unable to do so.
We have shown that it is possible to classify galaxies into AGN, 
passive galaxies and line emitting galaxies well without galaxy 
spectra, using solely colours and a training set of the order 
of 15000 galaxies.

We have discussed the prospects of speeding up redshift surveys with this 
and we conclude that with a reasonable training set of the order of 10000
galaxies one would be able to considerably speed up future surveys 
done with instruments such as FMOS and WFMOS. Furthermore, in this paper 
we have only taken advantage of the colour data;
however the methods described can accommodate very easily other information 
such as morphology or size.

\section*{Acknowledgements.}

We acknowledge Manda Banerji and Eduardo Cypriano for useful 
comments as well as Richard Ellis,
Peder Norberg, John Peacock and other members of one of the WFMOS
design study team.
FBA acknowledges the support of the Leverhulme Trust
via an Early Careers Fellowship. WAS and LSJ thanks the support of the 
Brazilian agencies FAPESP and CNPq.

\bibliography{}

\appendix

\section{Locally weighted regression}

Locally Weighted Regression (LWR) is a machine learning method 
capable of mapping complex, non-linear relations between variables \citep{LWR}.

Like any other global fits, 
LWR minimises a quantity related to the chi-square. 
The difference is that each data point of the training set has a 
corresponding weight, which depends on a given query point. Thus, 
the fitting parameters are valid only locally for that particular query point.
In this method any data available for the machine learning would 
be separated into a training set and a validation set.

In this method, the quantity to be minimised for each query point is given by

\begin{equation}
E = \sum_i w_i^2(y_i-f(\mathbf{x_i}))^2
\end{equation}

\noindent where $f$ is the linear function to be fitted for the weights 
for each point $w_i$ depend on the distance between the query point 
and the data point of the training set. The sum over i is performed for each 
point of the training set. Here the weights have been chosen to follow

\begin{equation}
w_i= {\rm exp}\, (- {\rm D}^2(\mathbf{x_i},\mathbf{x_q})/2K^2)
\end{equation}

\noindent where the function D is the Euclidean distance between the query 
point $x_q$ and the training data point $x_i$. The parameters $K$ is 
called the kernel width, related to the width of the Gaussian determining 
the weights. For reasonable values of $K$ only the data near the query 
point have significant contribution to the local fit. The optimal value 
for the parameter $K$ for a given training set is found by minimising 
the error found in a validation set. In the training stage many kernel values
are attempted and the one which produces best results for the validation 
set is chosen and fixed. Once K is chosen with the aid of the 
training and validation set, one is left to define how $f$ is chosen.

We choose to expand the fitting function f 

\begin{equation}
f(x)= \beta_1 t_1(x) +  \beta_2 t_2(x) + ... +  \beta_M t_M(x)
\end{equation}

\noindent where the functions $t_i$ are formed by linear combinations of
the inputs $x_i$; i.e. $t_1 = 1$, $t_2 = x_1$, $t_3 = x_1^2$, 
and so forth. This equation can be written as

\begin{equation}
f(x)= \mathbf{\beta}^T \mathbf{t}(x)
\end{equation}

\noindent where $t(x)$ is the vector of polynomial terms.
Here the weights are recomputed according the distance between points $x_k$
and the query point and the matrix $\beta$ is computed according to 

\begin{equation}
\mathbf{\beta} = (\mathbf{X}^T \mathbf{X})^{-1}\,\mathbf{X}^T y 
\end{equation}

\noindent where

\begin{equation}
(\mathbf{X}^T \mathbf{X})_{ij} = \sum_{k=1}^N  w_i^2 t_i(x_k)t_j(x_k)
\end{equation}

\noindent and

\begin{equation}
(\mathbf{X}^T y)_i = \sum_{k=1}^N  w_i^2 t_i(x_k)y_i
\end{equation}

\noindent hence we can solve for the best solution by using a Cholesky 
decomposition \citep{1992nrfa.book.....P}. Now that the $\mathbf{\beta}$
vector and the kernel are chosen we simply apply the relation using the 
testing set as the query points.

\section{Artificial neural networks}
\label{sec.anns}

\begin{figure}
\includegraphics[width=9.0cm,angle=0]{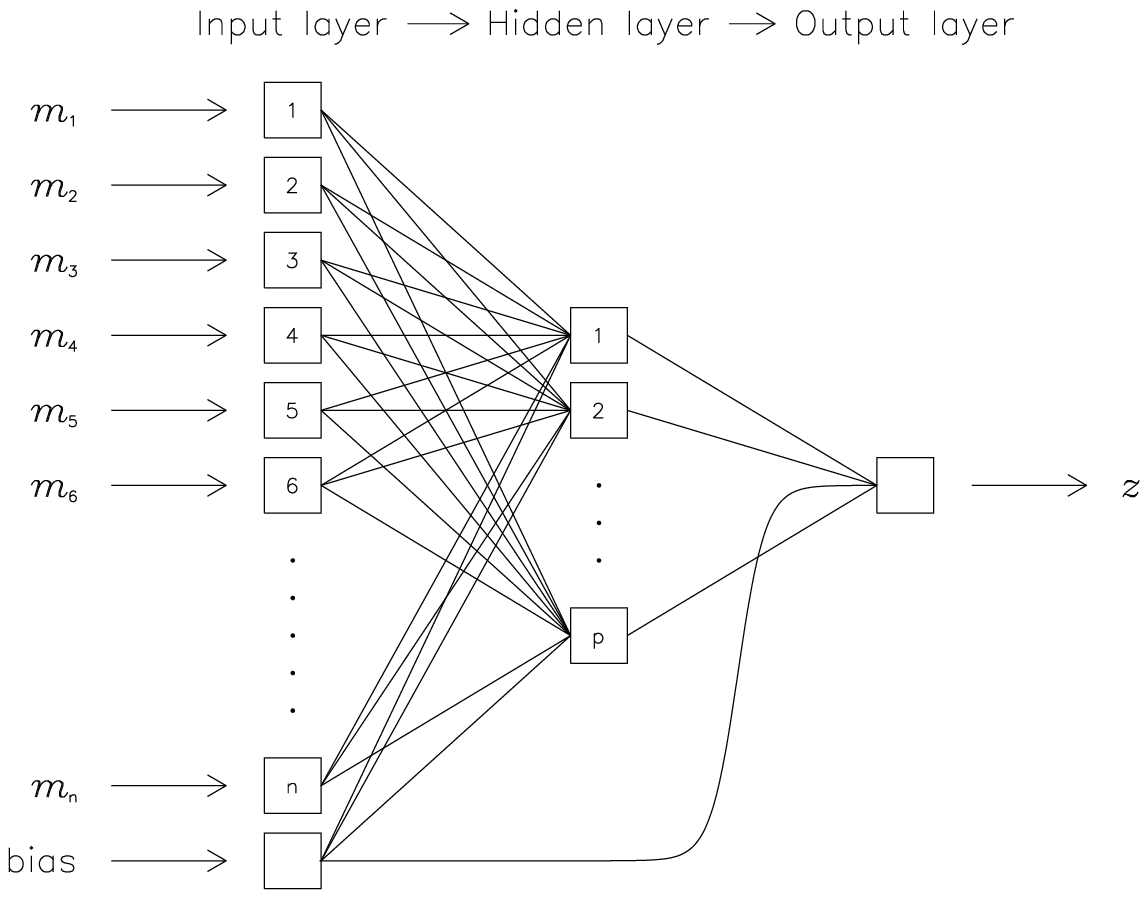}
\caption{\label{fig.photz.arch} A schematic diagram of a multi-layer
perceptron, as implemented by \textsc{ann}\emph{z}, with input nodes
taking, for example, magnitudes $m_i = -2.5\log_{10}f_i$ in various
filters, a single hidden layer, and a single output node.  
The architecture is $n$:$p$:1 in the notation
used.  Each connecting line carries a weight
$w_{ij}$. The bias node allows for an additive constant in the network
function defined at each node.  More complex networks can have
additional hidden layers and/or outputs. Here the equivalent width takes the
same role as the redshift.}
\end{figure}

We use a particular species of ANN known formally as a
\emph{multi-layer perceptron} (MLP). A MLP consists of a number of
layers of \emph{nodes} (Fig. \ref{fig.photz.arch}; see
e.g. \citealt{Bishop}, and references therein, for background). The
first layer contains the inputs, which in this paper 
are the magnitudes, $m_i$, of a galaxy
in a number of filters (for ease of notation we arrange these in a
vector $\mathbf{m} \equiv (m_1,m_2,...,m_n)$). The final layer
contains the outputs; here the
equivalent width or emission probability. 
Intervening layers are described
as \emph{hidden} and there is complete freedom over the number and
size of hidden layers used. The nodes in a given layer are connected
to all the nodes in adjacent layers. A particular network architecture
may be denoted by $N_{\mathrm{in}}$:$N_1$:$N_2$: $\ldots$
:$N_{\mathrm{out}}$ where $N_{\mathrm{in}}$ is the number of input
nodes, $N_1$ is the number of nodes in the first hidden layer, and so
on.  For example 9:6:1 takes 9 inputs, has 6 nodes in a single hidden
layer and gives a single output.

Each connection carries a weight, $w_{ij}$; these comprise the vector
of coefficients, $\mathbf{w}$, which are to be optimised. An
\emph{activation function}, $g_j(u_j)$, is defined at each node,
taking as its argument
\begin{equation}
u_j = \sum_i w_{ij}g_i(u_i), \label{eqn.act}
\end{equation}
where the sum is over all nodes $i$ sending connections to node
$j$. The activation functions are typically taken (in analogy to
biological neurons) to be sigmoid functions such as $g_j(u_j) = 1/[1 +
\exp{(-u_j)}]$, and we follow this approach here. An extra input node
-- the bias node -- is automatically included to allow for additive
constants in these functions.

For a particular input vector, the output vector of the network is
determined by progressing sequentially through the network layers,
from inputs to outputs, calculating the activation of each node (hence
this type of neural network is often referred to as a
\emph{feed-forward} network).

Given a suitable training set of galaxies for which we have both
photometry, $\mathbf{m}$, and here the Equivalent width,
$EW_\mathrm{train}$, the ANN is trained by minimising the \emph{cost
function}
\begin{equation}
E=\sum_k(EW_{\mathrm{out}}(\mathbf{w},\mathbf{m}_{k}) -
EW_{\mathrm{train},k})^2, \label{eqn.cost}
\end{equation}
with respect to the weights, $\mathbf{w}$, where
$EW_{\mathrm{out}}(\mathbf{w},\mathbf{m}_{k})$ is the network output
for the given input and weight vectors, and the sum is over the
galaxies in the training set. To ensure that the weights are
\emph{regularised} (i.e. that they do not become too large), an extra
quadratic cost term
\begin{equation}
E_\mathrm{w}=\beta\sum_{i,j}w_{ij}^2, \label{eqn.decay}
\end{equation}
is added to equation \ref{eqn.cost}. 

We use an iterative quasi-Newton method to perform this
minimisation. Details of the minimisation algorithm and regularisation
may be found in \citet{Bishop} and \citet[][
Appendices]{1996MNRAS.283..207L}.

After each training iteration, the cost function is also evaluated on
a separate \emph{validation} set. After a chosen number of training
iterations, training terminates and the final weights chosen for the
ANN are those from the iteration at which the cost function is minimal
on the \emph{validation} set. This is useful to avoid over-fitting to
the training set if the training set is small. The trained network may
then be presented with previously unseen input vectors, and the
outputs computed.

To implement the 
ANNs\footnote{http://zuserver2.star.ucl.ac.uk/$\sim$lahav/annz.html} code to our problem all that is needed
is to regard the output node as the equivalent width
instead of as the redshift.
In this work we found it optimal to work with the log(EW).

\end{document}